\RequirePackage{fix-cm}
\documentclass[smallextended]{svjour3}       % onecolumn (second format)
\usepackage[italian,english]{babel}
\smartqed  % flush right qed marks, e.g. at end of proof
\usepackage{amsmath}
\usepackage{amssymb}
\usepackage{lipsum}
\usepackage{dsfont}
\usepackage{multirow}
\usepackage{afterpage}
\usepackage{amsfonts}
\usepackage{bm}
\usepackage{babel}
\usepackage{hyperref}
\usepackage{graphicx}
\usepackage{color}
\usepackage{graphicx}
\begin{document}

\title{Characterizing nonbilocal correlation: A geometric perspective %\thanks{Grants or other notes
}


\author{R. Muthuganesan$^*$ \and   S. Balakrishnan 
\and  V. K. Chandrasekar }


\institute{\at Center for Nonlinear Science \& Engineering, School of Electrical \& Electronics Engineering, SASTRA Deemed University, Thanjavur, Tamil Nadu 613 401, India
             \email{rajendramuthu@gmail.com}           
             \and 
Department of Physics, School of Advanced Sciences, Vellore Institute of Technology, Vellore– 632014, India\at
 %             Tel.: +91-8760372825\\
              \email{physicsbalki@gmail.com}           %  \\
           \and
         Center for Nonlinear Science \& Engineering, School of Electrical \& Electronics Engineering,
SASTRA Deemed University, Thanjavur, Tamil Nadu 613 401, India\at
              \email{chandru@gmail.com}\\
              $^*$Corresponding Author.
}

\date{Received: date / Accepted: date}

\maketitle

\begin{abstract}
Exploiting the notion of measurement-induced nonlocality [Phys. Rev. Lett. {\bf 106}, 120401 (2011)], we introduce a new measure to quantify the nonbilocal correlation. We establish a simple relation between the nonlocal and nonbilocal measures for the arbitrary pure input states. Considering the mixed states as inputs, we derive two upper bounds of affinity-based nonbilocal measure. Finally,  we have studied the nonbilocality of a different combinations of input states.
\end{abstract}


\section{Introduction}
\label{intro}
Nonlocality, the most fundamental and intriguing feature of a composite quantum system, is a direct consequence of the superposition principle, which creates a distinction between the behavior of the quantum  and classical systems \cite{Nielsen2010}. Nonlocality is referred as "spooky-action-at-a-distance" by Einstein \cite{Einstein} and Schrodinger \cite{scho}.  Understanding this perplexing phenomenon in a simplest composite system, namely the bipartite system, is a fundamental issue and of practical importance in developing quantum technologies. In the realm of Bell's nonlocality \cite{Bell}, the presence of nonlocal character or entanglement is witnessed by the violation of Bell inequality. In the earliest quantum information theory era, it is believed that the entanglement is the complete manifestation of nonlocality of a quantum system. 

Since the inception of Werner's work \cite{Werner} and quantum discord \cite{Ollivier2001}, it has been debated whether the entanglement can manifest the nonlocal aspects of a quantum system or not. Buscemi showed that entanglement is considered as the complete manifestation of nonlocality. In other words, "all entangled quantum states are nonlocal" \cite{Buscemi}. On the other hand, Werner constructed a mixed state family, which admits the local hidden variable model even the state is entangled \cite{Werner}. Further, it is shown that the presence of noise or mixedness is responsible for the destroying nonlocal correlation between local constituents  of the composite system,  and hence some of the mixed entangled states behave locally \cite{Almeida}. Ollivier and Zurek introduced a measure called,  quantum discord  to quantify the quantum correlations  beyond entanglement. It can capture the correlation between the marginal states which cannot grasp by the entanglement \cite{Ollivier2001}. Recently, a new variant of nonlocality, called measurement-induced nonlocality (MIN) is introduced \cite{Luo2011}. MIN 
is based on the fact that local disturbance due to locally invariant von Neumann projective measurements on the marginal state can influence globally. MIN is quite different from the entanglement and violation of the Bell inequalities, more importantly it goes beyond entanglement.  

In quantum entanglement swapping experiment, the independence between the multi-sources 
induces the nonlocal behavior of probability distributions  and is called nonbilocal correlations \cite{Branciard2010,Branciard2012}. This kind of correlation is demonstrated and captured using nonlinear inequalities and one important class of these inequalities is the so-called binary-input-and-output bilocality inequality which is known as the bilocality inequality \cite{Branciard2010,Branciard2012}. In recent times, the considerable progress  has been made in this context \cite{Fritz2012,Fritz2016,Wood2015,Henson,Chaves2015,Tavakoli2014,Tavakoli2016a,Tavakoli2016b,Chaves2016,Rosset}. Gisin et al. have shown that pair of entangled states can violate the bilocality inequality, implying that tensorizing states may possess nonlocal correlation \cite{Gisin2017}.  

When two bipartite states with vanishing correlation are combined, it is always interesting to check that the tensorizing state possess any nonlocal advantage or not. It is shown recently that the combining two quantum systems exhibit better quantum advantages than the individual system. This is known as superactivation of nonlocality, symbolized as $0 + 0>0$  and it cannot occur in the classical world. The superactivation of nonlocality provides an answer for "can the state $\rho \otimes \rho$ be nonlocal if the $\rho$ is local". Recently, Palazuelos explained the superactivation of quantum nonlocality in the sense of violating certain Bell inequalities with an entangled bound state \cite{Palazuelos}. The same study is carried out in the context of tensor networks \cite{Cavalcanti2011,Cavalcanti2012}. Further, the superactivation was also considered for arbitrary entangled states by allowing local preprocessing on the tensor product of different quantum states $(\rho\otimes\rho')$ \cite{Masanes} and symbolized as $1 + 0>1$.   With our best knowledge, the quantification of this nonbilocal correlation is limited in the literature. 
 
 To quantify the nonbilocal correlation, we extend the notion of bipartite measurement-induced nonlocality (MIN) to a bilocal quantum system. This paper proposes a new version of the non-bilocal correlation measure. The relation between the nonlocal and non-bilocal correlation measure is established, and it is shown that non-bilocality is always greater than the nonlocal correlation. Further, the upper bound of the bilocal correlation measure is obtained for the arbitrary mixed input states. To validate the properties of nonbilocal measure, we study the proposed quantity for a few examples.  

This paper is organized as follows. In Sec. \ref{Sec2}, we review the notion of measurement-induced nonlocality and provide the definition of affinity based MIN.  In Sec. \ref{Sec3},  we introduce a nonbilocal measure based on the affinity induced metric and we derive the analytical formula of the measure when the input states are pure. Considering the input states are arbitrary mixed states, the upper bounds of bilocal correlation are presented in Sec. \ref{Sec4}.  In Sec. \ref{Sec5}, the proposed measure studied for a few examples.  Finally, in Sec. \ref{concl} we present the conclusion.

%%%%%%%%%%%%%%%%%%%%%%%%%%%%%%%%%%%%%%%%%%%%%%%%%%%%%%%%%%%%%%%%%%%%%%%%%%%%%%%%%%%%%%%%%%%%%%%%%%%%%%%%%%%%%%%%%%%%%%%%%%%%%%%%%%%%%%%%%%
\section{Measurement-induced nonlocality} 
\label{Sec2}
To capture the bipartite quantum correlation beyond entanglement, Luo and Fu introduced a new measure of  quantum correlation  called measurement-induced nonlocality (MIN) using locally invariant projective measurement. It is originally defined as maximal square of Hilbert-Schmidt norm of difference of pre- and post- measurement states and is defined as \cite{Luo2011}
\begin{equation}
 N(\rho ) =~^{\text{max}}_{\Pi ^{a}}\| \rho - \Pi ^{a}(\rho )\| ^{2},  \label{HS-MIN}
\end{equation}
where the maximization is taken over the von Neumann projective measurements on subsystem $a$, $\Pi^{a}(\rho) = \sum _{k} (\Pi ^{a}_{k} \otimes   \mathds{1} ^{b}) \rho (\Pi ^{a}_{k} \otimes    \mathds{1}^{b} )$, and $\Pi ^{a}= \{\Pi ^{a}_{k}\}= \{|k\rangle \langle k|\}$ being the projective measurements on the subsystem $a$, which do not change the marginal state $\rho^{a}$ locally i.e., $\Pi ^{a}(\rho^{a})=\rho ^{a}$. Here $\|\mathcal{O} \|=\sqrt{\text{Tr}\mathcal{O}^{\dagger}\mathcal{O}} $  is the   Hilbert-Schmidt norm of operator $\mathcal{O} $. The dual of this quantity is geometric discord (GD) of the given state $\rho$ and it is defined as \cite{Dakic,Luo2010pra}
\begin{equation}
 D(\rho ) =~^{\text{min}}_{\Pi ^{a}}\| \rho - \Pi ^{a}(\rho )\| ^{2}. 
\end{equation}
In general, if $\rho^{a}$ is nondegenerate, then the optimization is not required and the above measures are equal. This is more general than the conventionally mentioned quantum nonlocality related to Bell's version of nonlocality. Apart from the quantification of bipartite quantum correlation, this quantity provides a novel classification scheme for bipartite states, and is a useful resource quite different from entanglement. In particular, MIN is a more secured resource for cryptographic communication. Due to its geometric nature, MIN is easy to compute and realizable. Nevertheless, both the MIN and geometric discord are not useful quantifiers of quantum correlation due to the local ancilla problem, which is pointed out by Piani \cite{Piani2012}.  A natural way to circumvent this issue is to modify the definition of MIN using any contractive distance measure. One such distance measure between the states $\rho$ and $\sigma$ is defined as 
\begin{align}
d_{\mathcal{A}}(\rho,\sigma)=1-  \mathcal{A}(\rho,\sigma),
\end{align}
where $\mathcal{A}(\rho,\sigma)=\text{Tr}\left(\sqrt{\rho}\sqrt{\sigma}\right)$ is the affinity between the states. Analogous to fidelity \cite{Jozsa1994}, the affinity is also a measure of closeness between the states  $\rho$ and $\sigma$ \cite{Luo2004,Bhattacharyya} and shares all the properties of fidelity \cite{Jozsa1994}. Also, affinity is useful in the quantification of  nonclassical correlations \cite{Muthu2019,Muthu2020} and quantum coherence \cite{Muthu2021}. It is worth mentioning that the affinity between the states is realizable using the quantum circuit \cite{Ekert}. Due to its realization, affinity based measures may have a good impact in the  research of  quantum information. 

The affinity-based MIN is defined as \cite{Muthu2020}
\begin{align}
N^{\text{MIN}}_{\mathcal{A}}(\rho)=~^\text{{max}}_{ \Pi^{a}} ~  d_{\mathcal{A}}(\rho,\Pi^a(\rho))=1-~^\text{{min}}_{ \Pi^{a}} ~\text{Tr}\left(\sqrt{\rho}\Pi^a(\sqrt{\rho})\right).
\end{align}
Here also the optimization is taken over von Neumann projective measurements. It is worth reiterating  that the $N_{\mathcal{A}}(\rho)$ fixes the local ancilla problem using the multiplicative property of affinity. Hence, $N^{\text{MIN}}_{\mathcal{A}}(\rho)$ is a faithful quantifier of quantum correlation or quantumness of the system.  Further, we have shown that affinity based MIN is closely related to local quantum uncertainty \cite{Girolami2013LQU} and interferometric power of a quantum state \cite{Girolami2014}.  
% \begin{figure*}[!ht]
% \centering\includegraphics[width=0.4\linewidth]{page02new.jpg}
% \centering\includegraphics[width=0.5\linewidth]{page1.jpg}
% %\centering\includegraphics[width=0.4\linewidth]{Fig2c}
% %\centering\includegraphics[width=0.4\linewidth]{Fig2d}
% \caption{(color online) Schematic representation of nonbilocal correlation in tensorizing states (left). A scheme to measure nonbilocal correlation due to von Neumann projective measurements $\Pi^{bc}$(right). }
% \label{fig1}
% \end{figure*}
%%%%%%%%%%%%%%%%%%%%%%%%%%%%%%%%%%%%%%%%%%%%%%%%%%%%%%%%%%%%%%%%%%%%%%%%%%%%%%%%%%%%%%%%%%%%%%%%%%%%%%%%%%%%%%%%%%%%%%%%%%%%%%%%%%%%%%%%%%
\section{Nonbilocality Measure}\label{Sec3}
In this section, we introduce the notion and measure of nonbilocality using affinity. We consider two input states $\rho^{ab}$ (shared between $a$ and $b$) and $\rho^{cd}$ (shared between $c$ and $d$) in the separable composite finite-dimensional Hilbert space  $\mathcal{H}=\mathcal{H}_a\otimes\mathcal{H}_b\otimes\mathcal{H}_c\otimes\mathcal{H}_d$. Then nonbilocal measure is defined as 
\begin{align}
N_{\mathcal{A}}(\rho_{ab}\otimes\rho_{cd})=~^{\text{~max}}_{\Pi ^{bc}} ~d_{\mathcal{A}}(\rho_{ab}\otimes\rho_{cd}, \Pi^{bc}(\rho_{ab}\otimes\rho_{cd})) \nonumber \\
=1-~^{\text{min}}_{\Pi ^{bc}} \mathcal{A}(\rho_{ab}\otimes\rho_{cd}, \Pi^{bc}(\rho_{ab}\otimes\rho_{cd})),
\label{nonbilocal}
\end{align}
where maximization/minimization is taken over the von Neumann projective measurement $\Pi^{bc}=\{ \Pi^{bc}_k\} $, which leaves $\rho^{bc}=\text{Tr}_{ad}(\rho_{ab}\otimes\rho_{cd})$  invariant locally,  $d_{\mathcal{A}}(\cdot, \cdot)$ is affinity-induced metric and the post-measurement state is $\Pi^{bc}(\sqrt{\rho_{ab}\otimes\rho_{cd}})=\sum_{k,l}(\mathds{1}^a \otimes \Pi^{bc}_{kl} \otimes\mathds{1}^d)\sqrt{\rho_{ab}\otimes\rho_{cd}}(\mathds{1}^a \otimes \Pi^{bc}_{kl} \otimes\mathds{1}^d)$. Here $\rho^b=\sum_i\lambda_i| i_b\rangle \langle i_b| $ and $\rho^c=\sum_j\lambda_j| j_c\rangle \langle j_c| $ are the marginal states of $\rho_{bc}$, if any one of the states is nondegenerate, then the measurement takes the form $\Pi^{bc}=\{\Pi^{b}\otimes\Pi^{c} \} $.

It is worth mentioning that above defined quantity satisfies all the necessary axioms of a quantum correlation measure. Like MIN, the nonbilocal measure is also a useful resource for cryptographic communication. Next, we demonstrate some interesting properties of nonbilocality measure given by Eq. (\ref{nonbilocal}).
\begin{enumerate}
\item[(i)] $N_{\mathcal{A}}(\rho_{ab}\otimes\rho_{cd})$ is nonnegative, 

\item[(ii)] $N_{\mathcal{A}}(\rho_{ab}\otimes\rho_{cd})= 0$ for any product states $\rho_{ab}=\rho^a\otimes\rho^b$ and $\rho_{cd}=\rho^c\otimes\rho^d$. Further, the nonbilocal measure also vanishes for classical-quantum state $\rho_{ab}=\sum_i\rho^a_i\otimes p_i| i_b\rangle \langle i_b|$ and $\rho_{cd}=\sum_j q_j| j_c\rangle \langle j_c|\otimes \rho^d_j$.
\item[(iii)]   $N_{\mathcal{A}}(\rho_{ab}\otimes\rho_{cd})$  is locally unitary invariant in the sense that 
\begin{align}
N_{\mathcal{A}}((U_{ab}\otimes U_{cd})\rho_{ab}\otimes\rho_{cd}(U_{ab}\otimes U_{cd})^{\dagger})=N_{\mathcal{A}}(\rho_{ab}\otimes\rho_{cd}),
\end{align}
where $U_{ab}=U_a\otimes U_b$ and $U_{cd}=U_c\otimes U_d$ are the local unitary operators.
\item[(iv)] $N_{\mathcal{A}}(\rho_{ab}\otimes\rho_{cd})$  is positive, at least any one of the input states is entangled. 

\item[(v)] Although $N_{\mathcal{A}}(\rho_{ab})=N_{\mathcal{A}}(\rho_{cd})=0$, nevertheless $N_{\mathcal{A}}(\rho_{ab}\otimes\rho_{cd})> 0$.
\end{enumerate}
Since the properties (i) - (v) are inherited directly from the properties of affinity-based MIN, we omit the proving process. Next, we establish a simple relation between the nonbilocal measure and MIN (nonlocal).

\textit{Theorem 1: The nonbilocal measure and affinity-based MIN are connected as
\begin{align}
N_{\mathcal{A}}(\rho_{ba}\otimes\rho_{ab}) \geq N_{\mathcal{A}}^{\text{MIN}}(\rho).
\end{align}
} 
Proof: To prove this, first we recall the definition of affinity-based nonbilocal measure,
\begin{align}
N_{\mathcal{A}}(\rho_{ba}\otimes\rho_{ab}) =&~^{\text{~max}}_{\Pi ^{aa}} ~d_{\mathcal{A}}(\rho_{ba}\otimes\rho_{ab}, \Pi^{aa}(\rho_{ba}\otimes\rho_{ab})), \nonumber \\
=&1-~^{\text{min}}_{\Pi ^{aa}} \mathcal{A}(\rho_{ba}\otimes\rho_{ab}, \Pi^{aa}(\rho_{ba}\otimes\rho_{ab})), \nonumber \\
\geq& ~^{\text{~max}}_{\Pi ^{a}} ~d_{\mathcal{A}}(\rho_{ba}\otimes\rho_{ab}, (\Pi^{a}\otimes\Pi^{a})(\rho_{ba}\otimes\rho_{ab})), \nonumber \\
=&1-~^{\text{~min}}_{\Pi ^{a}} \text{Tr}(\sqrt{\rho_{ab}}\otimes\sqrt{\rho_{ab}})(\Pi^a(\sqrt{\rho_{ab}})\otimes\Pi^a(\sqrt{\rho_{ab}})), \nonumber \\
=&1-~^{\text{~min}}_{\Pi ^{a}} \text{Tr}(\sqrt{\rho_{ab}} \Pi^a(\sqrt{\rho_{ab}}))^2, \nonumber \\
\geq& 1-~^{\text{~min}}_{\Pi ^{a}} \text{Tr}(\sqrt{\rho_{ab}} \Pi^a(\sqrt{\rho_{ab}})),  \nonumber \\
=&N_{\mathcal{A}}^{MIN}(\rho), \nonumber
\end{align} 
where the first inequality follows from the fact that $\Pi^a\otimes \Pi^a$ is not necessarily optimal and the second inequality is due to the square of the affinity which is always equal to or less than the affinity. Hence the theorem is proved. The above theorem provides a closer connection between the nonbilocal and nonlocal measures and implying that the nonbilocal measure is always greater than MIN. Next, we compute the nonbilocal measure for pure input states.

\textit{Theorem 2: Let $|\Psi_{ab}\rangle=\sum_i\sqrt{s_i}|i_a  i_b \rangle$  and $|\Psi_{ab}\rangle =\sum_j\sqrt{r_j}|j_c j_d\rangle$ are the pure input states, then 
\begin{align}
N_{\mathcal{A}}(|\Psi_{ab}\rangle\otimes|\Psi_{cd}\rangle)=1-\sum_{i,j}s_i^4r_j^4,
\end{align}
where $s_i$ and $r_j$ are Schmidt coefficients of $|\Psi_{ab}\rangle$ and $|\Psi_{cd}\rangle$ respectively.} 

The proof of the theorem is given in the appendix. 

\section{Nonbilocal correlation for mixed states}
\label{Sec4}
To compute nonbilocality of any arbitrary mixed input states, first, we recall some basic notation in the operator Hilbert space. Let $\mathcal{L}(\mathcal{H}_{\alpha})$ be the Hilbert space of linear operators on $\mathcal{H}_{\alpha} (\alpha=a,b,c,d)$ with the inner product $\langle X|Y\rangle=\text{Tr}X^{\dagger}Y $. An arbitrary  $m\times n$ dimensional bipartite state can be written as
\begin{align}
\sqrt{\rho_{ab}}=\sum_{i,j} \lambda^{ab}_{ij} X_i\otimes Y_j, \nonumber
%\label{stateab}
\end{align}
where $\{X_i : i=0,1, \cdots, m^2-1 \}$ and  $\{Y_j : j=0,1, \cdots, n^2-1 \}$ are the orthonormal operator bases of the subsystem $a$ and $b$ respectively and satisfies the relation $\text{Tr}X_kX_l=\delta_{kl}$, and $\lambda^{ab}_{ij}=\text{Tr}(\sqrt{\rho_{ab}}~X_i\otimes Y_j)$ are real elements of matrix $\Lambda_{ab}$. Similarly, one can define the orthonormal operator bases  as $\{P_k : k=0,1, \cdots, u^2-1 \} ~\text{and~} \{Q_l : l=0,1, \cdots, v^2-1 \}$ for another input state $\rho_{cd}$ with $u$ and $v$ are the dimensions of the marginal systems $c$ and $d$ respectively.  Then the state $\rho_{cd}$ is defined as 
\begin{align}
\sqrt{\rho_{cd}}=\sum_{k,l} \lambda^{cd}_{kl} P_k\otimes Q_l,  \nonumber
%\label{state2}
\end{align}
where $\lambda^{cd}_{kl}=\text{Tr}(\sqrt{\rho_{cd}}~P_k\otimes Q_l)$ are the matrix elements of matrix $\Lambda_{cd}$. Then the bilocal state is written as 
\begin{align}
\sqrt{\rho_{ab}\otimes \rho_{cd}}=\sum_{i,j} \sum_{k,l} \lambda^{ab}_{ij}\lambda^{cd}_{kl}  X_i\otimes Y_j\otimes P_k\otimes Q_l.
\label{bilocal}
\end{align}

\textit{Theorem 3: For any arbitrary bilocal input states represented in Eq. (\ref{bilocal}), the upper bound of nonbilocal measure is
\begin{align}
N_{\mathcal{A}}(\rho_{ab}\otimes\rho_{cd})\leq 1-\sum_{s=1}^{nu}\mu_{s},
\end{align}
where $\mu_{s}$ are the eigenvalues of the matrix $\Lambda_{ab,cd}\Lambda^t_{ab,cd}$ arranged in increasing order and $t$ denotes the transpose of a matrix. 
}

To prove the theorem, first, we compute the post-measured state. Here, the measurement operators are $\Pi^{bc}=\{ \mathds{1}^a\otimes\Pi^{bc}_h\otimes\mathds{1}^d\} $ and we have
\begin{align}
\Pi^{bc}(\sqrt{\rho_{ab}\otimes \rho_{cd}})=&\sum_h \sum_{ijkl}  \lambda^{ab}_{ij}\lambda^{cd}_{kl} X_i\otimes\Pi^{bc}_h(Y_j\otimes P_k)\Pi^{bc}_h\otimes Q_l \nonumber \\
=& \sum_h \sum_{ijj'kk'l}  \lambda^{ab}_{ij}\lambda^{cd}_{kl} \gamma_{hjk} \gamma_{hj'k'} X_i\otimes Y_{j'}\otimes P_{k'}\otimes Q_l,
\end{align}
where $\gamma_{hjk}=\text{Tr}\Pi^{bc}_h(Y_j\otimes P_k)$ are the elements of matrix $\Gamma$. After the straight forward calculation the affinity between pre- and post-measurement states is 
\begin{align}
\mathcal{A}(\sqrt{\rho_{ab}\otimes \rho_{cd}},\Pi^{bc}(\sqrt{\rho_{ab}\otimes \rho_{cd}}))=&\sum_h \sum_{ijj'kk'l} ~ \lambda^{ab}_{ij}\lambda^{cd}_{kl} \gamma_{hjk} \gamma_{hj'k'}\lambda^{ab}_{ij'}\lambda^{cd}_{k'l} \nonumber \\
=& \Gamma \Lambda_{ab,cd}\Lambda^t_{ab,cd} \Gamma^t,
\end{align}
where $\Gamma$ is $nu\times n^2 u^2$   dimensional matrix. Then, 
\begin{align}
N_{\mathcal{A}}(\rho_{ab}\otimes\rho_{cd})=1-~^{\text{min}}_{~\Gamma}~\Gamma \Lambda_{ab,cd}\Lambda^t_{ab,cd} \Gamma^t\leq 1-\sum_{s=1}^{nk}\mu_{s},
\end{align}
where $\mu_{s}$ are the eigenvalues of the matrix $\Lambda_{ab,cd}\Lambda^t_{ab,cd}$ listed in increasing order. Hence the theorem is proved.

\textit{Theorem 4: If the marginal state $\rho^b$ is nondegenerate,  the nonbilocal measure $N_{\mathcal{A}}(\rho_{ab}\otimes\rho_{cd})$  due to the measurement $\Pi^{bc}$ has the upper bound as 
\begin{align}
N_{\mathcal{A}}(\rho_{ab}\otimes\rho_{cd})\leq 1- {\mathcal{A}}(\rho_{ab}, \Pi^b(\rho_{ab}) \times (\sum_{\tau=1}^{u}\mu_{\tau}),
\end{align}
where $\mu_{\tau}$ are the eigenvalues of matrix $\Lambda_{cd}\Lambda^t_{cd}$ arranged in an increasing order and $\mathcal{A}(\rho_{ab}, \Pi^b(\rho_{ab}))$ is the affinity between the state $\rho_{ab}$ and post-measurement state $\Pi^{b}(\sqrt{\rho_{ab}})$. 
}

To show this, we recall that if the marginal state is nondegenerate, then the optimization is not required, Here the state $\rho^b$ is nondegenerate and the optimization is taken over $\Pi^c$ alone. The measurement operator is defined as $\Pi^b\otimes\Pi^c=\{ \Pi^b_j\otimes\Pi^c_k\}=\{| j_b\rangle \langle j_b| \otimes \Pi^c_k\} $. The nonbilocality measure is 
\begin{align}
N_{\mathcal{A}}(\rho_{ab}\otimes\rho_{cd})=&1-~^{\text{min}}_{\Pi ^{bc}} ~{\mathcal{A}}(\rho_{ab}\otimes\rho_{cd},\Pi ^{bc}(\rho_{ab}\otimes\rho_{cd})) \nonumber \\
=& 1-~^{\text{min}}_{\Pi ^{bc}} ~\text{Tr}\sqrt{\rho_{ab}\otimes\rho_{cd}}\cdot\Pi ^{bc}\sqrt{(\rho_{ab}\otimes\rho_{cd})} \nonumber \\
=& 1-~^{\text{min}}_{\Pi ^{c}}~\text{Tr}\sqrt{\rho_{ab}\otimes\rho_{cd}}\cdot(\Pi ^{b}\otimes\Pi ^{c})(\sqrt{\rho_{ab}\otimes\rho_{cd}}) \nonumber \\
=& 1-\text{Tr}\sqrt{\rho_{ab}}\Pi^{b}(\sqrt{\rho_{ab}})\cdot ~^{\text{min}}_{\Pi ^{c}}~\text{Tr}~\sqrt{\rho_{cd}}\Pi ^{c}(\sqrt{\rho_{cd}}),
\end{align}
where the quantity $\text{Tr}\sqrt{\rho_{ab}}\Pi^{b}(\sqrt{\rho_{ab}})$ is the affinity between the state $\rho_{ab}$ and post-measurement state $\Pi^{b}(\sqrt{\rho_{ab}})$. Following the optimization procedure given in \cite{Muthu2020},  we write the second quantity as 
\begin{align}
~^{\text{min}}_{~\Pi ^{c}}~\text{Tr}~\sqrt{\rho_{cd}}\Pi ^{c}(\sqrt{\rho_{cd}})=~^{\text{min}}_{~C}~\text{Tr}~C\Lambda_{cd}\Lambda_{cd}^tC^t.
\end{align}
 Then, we have
\begin{align}
N_{\mathcal{A}}(\rho_{ab}\otimes\rho_{cd})=&1-\mathcal{A}(\rho_{ab},\Pi^{b}(\sqrt{\rho_{ab}}))~^{\text{min}}_{~C}~\text{Tr}~C\Lambda_{cd}\Lambda_{cd}^tC^t \nonumber \\
\leq & 1- \mathcal{A}(\rho_{ab}, \Pi^b(\rho_{ab})) \times \sum_{\tau=1}^{u}\mu_{\tau},
\end{align}
where $\mu_{\tau}$ are the eigenvalues of matrix $\Lambda_{cd}\Lambda^t_{cd}$ arranged in an increasing order.

\textit{Theorem 5: If  the marginal states $\rho^b$ and $\rho^c$ are nondegenerate and the dimension of the  $\rho^c$ is $u=2$, then the closed formula of nonbilocal measure $N_{\mathcal{A}}(\rho_{ab}\otimes\rho_{cd})$ is expressed as  
\begin{align}
N_{\mathcal{A}}(\rho_{ab}\otimes\rho_{cd})=1- \mathcal{A}(\rho_{ab}, \Pi^b(\rho_{ab}) ) \times  \| \boldsymbol\lambda _{cd} \| +\lambda_{min},
\end{align}
where $ \boldsymbol\lambda_{cd}=(\lambda^{cd}_{00},\lambda^{cd}_{01}, \cdots, \lambda^{cd}_{0(v^2-1)})$ is a $v^2$ dimensional row vector, and $\Lambda=((\lambda^{cd})_{kl})_{k=1,2,3;~l=0,1, \cdots v^2-1}$ is a $3\times v^2$ dimensional matrix and $\lambda_{min}$ is the least eigenvalues of $\Lambda\Lambda^t$}

Using the completeness relation $\sum_k\Pi_k^c=\mathds{1}^c$, we show that $c_{0k}=-c_{1k} (k=1,2,3)$, then $\sum_k c_{0k}^2=1$. Then the vector $\mathbf{c}=\sqrt{2}(c_{0k}~c_{0k}~c_{0k})$ with $|\mathbf{c}|=1 $. Now, 
\begin{align}
C=\frac{1}{\sqrt{2}}
\begin{pmatrix}
1 & \mathbf{c}\\
1 & -\mathbf{c}
\end{pmatrix},
\end{align}
and 
\begin{align}
\Lambda_{cd}=
\begin{pmatrix}
\boldsymbol\lambda_{cd}\\
\Lambda
\end{pmatrix},
\end{align}
where $\boldsymbol\lambda_{cd}=(\lambda^{cd}_{00},\lambda^{cd}_{01}, \cdots, \lambda^{cd}_{0(v^2-1)})$ is a $v^2$ dimensional row vector, and $\Lambda=((\lambda^{cd})_{kl})_{k=1,2,3;~l=0,1, \cdots, v^2-1}$ is a $3\times v^2$ dimensional matrix. We have 
\begin{align}
^{\text{min}}_{~C}~\text{Tr}C\Lambda_{cd}\Lambda_{cd}^tC^t=\| \boldsymbol\lambda_{cd} \| +\lambda_{\text{min}},
\end{align}
where $\lambda_{min}$ is the least eigenvalue of matrix $\mathbf{c} RR^t \mathbf{c}^t$.
Then we have computed the closed formula of nonbilocal measure 
\begin{align}
N_{\mathcal{A}}(\rho_{ab}\otimes\rho_{cd})=~1-\mathcal{A}(\rho_{ab},\Pi^{b}(\sqrt{\rho_{ab}}))\times(\| \boldsymbol\lambda_{cd} \| +\lambda_{\text{min}})
\end{align}
to complete the proof.
%%%%%%%%%%%%%%%%%%%%%%%%%%%%%%%%%%%%%%%%%%%%%%%%%%%%%%%%

\section{Illustrations}
\label{Sec5}
In this section, we compute the affinity-based measurement-induced nonbilocality for some input states. 

\textit{Example 1:} Let $|\Psi_{ab}\rangle=|00\rangle $ and $|\Psi_{cd}\rangle=(|00\rangle+ |11\rangle)/\sqrt{2}$ are the two input states. 
According to Theorem. 2, the nonbilocal measure is 
\begin{align}
N_{\mathcal{A}}(|\Psi_{ab}\rangle\otimes|\Psi_{cd}\rangle)=1-\sum_{i,j}s_i^4r_j^4. 
\end{align}
The Schmidt coefficients for $|\Psi_{ab}\rangle$ are 0 and 1. Similarly, $|\Psi_{ab}\rangle$ has the Schmidt coefficients $1/\sqrt{2}$ and $1/\sqrt{2}$. Then, $N_{\mathcal{A}}(|\Psi_{ab}\rangle\otimes|\Psi_{cd}\rangle)=0.5$. The above example validate the property (iv) of the $N_{\mathcal{A}}(\rho_{ab}\otimes\rho_{cd})$.

\textit{Example 2:} The input state is $|\Psi_{ab}\rangle\otimes|\Psi_{cd}\rangle=(|00\rangle+|11\rangle)/\sqrt{2}\otimes(|00\rangle+|11\rangle)/\sqrt{2}$. Then, 
\begin{align}
N_{\mathcal{A}}(|\Psi_{ab}\rangle\otimes|\Psi_{cd}\rangle)=1-4 \times \frac{1}{4} \times \frac{1}{4} =\frac{3}{4}.
\end{align}

\textit{Example 3:} Next, we consider the combination of maximally entangled state as 
\begin{align}
\rho_{ab}=\frac{1}{3}(|\Psi^+\rangle\langle\Psi^+|+|\Psi^-\rangle\langle\Psi^-|+|\Phi^+\rangle\langle\Phi^+|), \nonumber
\end{align}
where $|\Psi^\pm\rangle=(|00\rangle\pm|11\rangle)/\sqrt{2} ~~\text{and}~~ |\Phi^\pm\rangle=(|01\rangle\pm|10\rangle)/\sqrt{2}$. In straight forward, the square root of $\rho_{ab}$ is 
\begin{align}
\sqrt{\rho_{ab}}=\frac{1}{\sqrt{3}}(|\Psi^+\rangle\langle\Psi^+|+|\Psi^-\rangle\langle\Psi^-|+|\Phi^+\rangle\langle\Phi^+|). \nonumber
\end{align}
The Bloch form of $\sqrt{\rho_{ab}}$ can be written as 
\begin{align}
\sqrt{\rho_{ab}}=\frac{1}{4}\left[\sqrt{3}(\mathds{1}\otimes\mathds{1}) +\frac{1}{\sqrt{3}}(\sigma_x\otimes\sigma_x)+\frac{1}{\sqrt{3}}(\sigma_y\otimes\sigma_y)+\frac{1}{\sqrt{3}}(\sigma_z\otimes\sigma_z)\right], \nonumber
\end{align}
where $\sigma_i$ are Pauli's spin matrices. Here, the eigenprojective measurements are $\Pi^{bc}=\{|\Psi^+\rangle\langle \Psi^+|, |\Psi^-\rangle\langle \Psi^-|,\\ |\Phi^+\rangle\langle \Phi^+|,|\Phi^-\rangle\langle \Phi^-| \} $. Then, the nonbilocal measure is computed as 
\begin{align}
N(\rho_{ba}\otimes\rho_{ab})=1-~^{\text{min}}_{~\Gamma}~\Gamma \Lambda_{ab,cd}\Lambda^t_{ab,cd} \Gamma^t\geq  1-\frac{7}{12}=\frac{5}{12}.
\end{align}
The affinity-based MIN is $ N_{\mathcal{A}}^{\text{MIN}}(\rho)=1/6< N(\rho_{ba}\otimes\rho_{ab})$, resulting the consequence of Theorem. 1.

\textit{Example 4:} In this case, the input states are
\begin{align}
\rho_{ab}=\rho_{cd}=\frac{1}{2}|0\rangle \langle 0|\otimes |0\rangle \langle 0|+\frac{1}{2}|1\rangle \langle 1|\otimes |1\rangle \langle 1|
\end{align}
shared between $a$ and $b$. The MIN of the above state is zero. We obtain 
\begin{align}
\Lambda_{ab}=\Lambda_{cd}=
\begin{pmatrix}
\frac{1}{\sqrt{2}} & 0 & 0 & 0 \\
0 & 0 & 0 & 0 \\
0 & 0 & 0 & 0 \\
0 & 0 & 0 & \frac{1}{\sqrt{2}} 
\end{pmatrix}.
\end{align}
Here we choose the optimal von Neumann measurements as $\{ \Pi^{bc}\}=\{H^{\otimes 2}|ij \rangle \langle ij| H^{\otimes2} \}  $ with $i,j=0,1$ and $H$ is  
\begin{align}
H=\frac{1}{\sqrt{2}}
\begin{pmatrix}
1 & 1\\
1 & -1
\end{pmatrix},
\end{align}
the popular single-qubit Hadamard gate. After the straight forward calculation, we have
\begin{align}
N(\rho_{ba}\otimes\rho_{ab})=1-~^{\text{min}}_{~\Gamma}~\Gamma \Lambda_{ab,cd}\Lambda^t_{ab,cd} \Gamma^t =\frac{3}{4}.
\end{align}
%%%%%%%%%%%%%%%%%%%%%%%%%%%%%%%%%%%
% \section{Interferometric Power}
% Here, we interpret the operational meaning of affinity-based nonbilocal measure in terms of interferometric power (IP). The IP quantifies the  discord-like quantum correlation with respect to the observable $H$. According to Girolami et al., the IP for tensorizing state is defined as  
% \begin{align}
%   IP(\rho_{ab}\otimes\rho_{cd})=~^\text{{~~~~~min}}_{ \mathds{1}\otimes|hk \rangle \langle hk |\otimes \mathds{1} } ~\mathcal{I}(\rho_{ab}\otimes\rho_{cd},\mathds{1}^a\otimes H^{bc} \otimes \mathds{1}^d)
% \end{align}
% where $\mathcal{I}$ is quantum Fisher information defined via symmetric logarithm derivatives and the minimum is intended over all fixed observable $H$.
%
%

%%%%%%%%%%%%%%%%%%%%%%%%%%%%%%%%%%%%%%%%%%%%%%%%%%%%%%%%%%%%%%%%%
%%%%%%%%%%%%%%%%%%%%%%%%%%%%%%%%%%%%%%%%%%%%%%%%%%%%%%%%%%%%%%%%%%%%%%%%%%%%%%%%%%%%%%%%%%%%%%%%%%
\section{Conclusion}
\label{concl}
Nonlocality is often related to the entanglement or violation of Bell's inequality. In this article, we have employed the nonlocality in different notion, namely measurement-induced nonlocality (MIN). Extending the definition of affinity-based MIN, we have introduced a new variant of quantifier to quantify the nonlocal correlation contained in tensorizing a local state with itself called nonbilocal correlation and also demonstrated the computational properties of the proposed measure.  A closer connection between the affinity-based measurement-induced nonlcality and measurement-induced nonbilcality is also derived. An analytical formula of nonbilocal measure is derived when the input states are pure. Moreover, two upper bounds of nonbilocal measure are also obtained for mixed input state. For illustration, we have studied the nonbilocality for different input states.

Like MIN, the nonbilocality measure may also useful resource for remote state preparation, quantum dense coding and cryptographic communication and hence, the proposed nonbilocal measure provides more insight into quantum information theory.

%\begin{figure}
  %\includegraphics{example.eps}
%\end{figure}
%\begin{figure*}
 % \includegraphics[width=0.75\textwidth]{example.eps}
%\end{figure*}

\noindent

\section*{Appendix}
%\begin{proof*}
Let $|\Psi_{ab}\rangle=\sum_is_i|i_ai_b\rangle$ and $|\Psi_{cd}\rangle=\sum_jr_j|j_cj_d\rangle$ are the two pure input states with $s_i$ and $r_j$ are the respective Schmidt coefficients of input states.  

Noting that 
\begin{equation}
\begin{aligned}
\rho_{ab}\otimes \rho_{cd}&=|\Psi_{ab}\rangle\langle\Psi_{ab}|\otimes|\Psi_{cd}\rangle\langle\Psi_{cd}|\\
&=\sum_{ii^{'}jj^{'}}s_is_{i^{'}}r_jr_{j^{'}}|i_a\rangle\langle i^{'}_a|\otimes |i_b\rangle\langle i^{'}_b|\otimes |j_c\rangle\langle j^{'}_c|\otimes |j_d\rangle\langle j^{'}_d|.
\end{aligned}
\end{equation}
One can compute the marginal state
\begin{equation}
\begin{aligned}
\rho^{bc}&=\text{Tr}_{ad}(|\Psi_{ab}\rangle\langle\Psi_{ab}|\otimes|\Psi_{cd}\rangle\langle\Psi_{cd}|)=\sum_{ij}s_i^2r_j^2|i_bj_c\rangle\langle i_bj_c|.
\label{marginal}
\end{aligned}
\end{equation}
For pure state $\sqrt{\rho}=\rho$. From the above equation, the post-measurement state $\Pi^{bc}(\sqrt{\rho_{ab}\otimes \rho_{cd}})$ can be rewritten as
\begin{equation*}
\begin{aligned}
&\Pi^{bc}(\sqrt{\rho_{ab}\otimes \rho_{cd}})=\Pi^{bc}(\rho_{ab}\otimes \rho_{cd})\\
=&\sum_{hk}(\mathds{1}^{a}\otimes\Pi_{hk}^{bc}\otimes \mathds{1}^{d})(|\Psi_{ab}\rangle\langle\Psi_{ab}|\otimes|\Psi_{cd}\rangle\langle\Psi_{cd}|)(\mathds{1}^{a}\otimes\Pi_{hk}^{BC}\otimes \mathds{1}^{d})\\
=&\sum_{hk}(\mathds{1}^{a}\otimes\Pi_{hk}^{bc}\otimes \mathds{1}^{d})(\sum_{ii^{'}jj^{'}}s_is_{i^{'}}r_jr_{j^{'}}|i_a\rangle\langle i^{'}_a|\otimes |i_b\rangle\langle i^{'}_b|\otimes |j_c\rangle\langle j^{'}_c|\otimes|j_d\rangle\langle j^{'}_d|)(\mathds{1}^{a}\otimes\Pi_{hk}^{bc}\otimes \mathds{1}^{d})\\
=&\sum_{hk}\sum_{ii^{'}jj^{'}}s_is_{i^{'}}r_jr_{j^{'}}|i_a\rangle\langle i^{'}_a|\otimes \Pi^{bc}_{hk}|i_b j_{c}\rangle\langle i^{'}_b j^{'}_c|\Pi^{bc}_{hk}\otimes |j_d\rangle\langle j^{'}_d|\\
=&\sum_{hk}\sum_{ii^{'}jj^{'}}s_is_{i^{'}}r_jr_{j^{'}}|i_a\rangle\langle i^{'}_a|\otimes U|h_{b}k_{c}\rangle\langle h_{b}k_{c}|U^{\dag}|i_b j_{c}\rangle\langle i^{'}_b j^{'}_c|U|h_{b}k_{c}\rangle\langle h_{b}k_{c}|U^{\dag}\otimes |j_d\rangle\langle j^{'}_d|.
%=&\sum_{ef}\sum_{ii^{'}jj^{'}}\lambda_i\lambda_{i^{'}}\mu_j\mu_{j^{'}}\langle e_{B}f_{C}|U^{\dag}|i_B j_{C}\rangle\langle i^{'}_B j^{'}_C|U|e_{B}f_{C}\rangle|i_A\rangle\langle i^{'}_A|\otimes U|e_{B}f_{C}\rangle\langle e_{B}f_{C}|U^{\dag}\otimes |j_D\rangle\langle j^{'}_D|,
\end{aligned}
\end{equation*}
Here the von Neumann projective measurement is expressed as 
%\begin{equation}
\begin{align}
\Pi^{bc}=\{\Pi^{bc}_{hk}\equiv U|h_{b}k_{c}\rangle\langle h_{b}k_{c}|U^{\dag}\}
\label{measure}
\end{align}
%\end{equation}
Consequently,
\begin{equation*}
\begin{aligned}
&\sqrt{\rho_{ab}\otimes \rho_{cd}}\Pi^{bc}(\sqrt{\rho_{ab}\otimes \rho_{cd}})\\
=&(\sum_{ii^{'}jj^{'}}s_is_{i^{'}}r_jr_{j^{'}}|i_a\rangle\langle i^{'}_a|\otimes |i_b\rangle\langle i^{'}_b|\otimes |j_c\rangle\langle j^{'}_c|\otimes |j_d\rangle\langle j^{'}_d|)(\sum_{hk}\sum_{uu^{'}vv^{'}}s_us_{u^{'}}r_vr_{v^{'}}\\
&\langle h_{b}k_{c}|U^{\dag}|u_b v_{c}\rangle\langle u^{'}_b v^{'}_c|U|h_{b}k_{c}\rangle|u_a\rangle\langle u^{'}_a|\otimes U|h_{b}k_{c}\rangle\langle h_{b}f_{c}|U^{\dag}\otimes |v_d\rangle\langle v^{'}_d|)\\
=&\sum_{ii^{'}jj^{'}}\sum_{hk}\sum_{uu^{'}vv^{'}}s_is_{i^{'}}r_jr_{j^{'}}s_us_{u^{'}}r_vr_{v^{'}}\langle e_{B}f_{C}|U^{\dag}|u_B v_{C}\rangle\langle u^{'}_b v^{'}_c|U|h_{b}k_{f}\rangle|i_a\rangle\\
&\langle i^{'}_a|u_a\rangle\langle u^{'}_a|\otimes|i_b j_c\rangle\langle i^{'}_b j^{'}_c|U|h_{b}k_{c}\rangle\langle h_{b}k_{c}|U^{\dag}\otimes |j_d\rangle\langle j^{'}_d|v_d\rangle\langle v^{'}_d|.
\end{aligned}
\end{equation*}
Then the affinity between the pre- and post-measurement state is computed as 
\begin{equation*}
\begin{aligned}
\mathcal{A}(\rho_{ab}\otimes \rho_{cd},\Pi^{bc}(\rho_{ab}\otimes \rho_{cd}))=&\text{Tr}\sqrt{\rho_{ab}\otimes \rho_{cd}}\Pi^{bc}(\sqrt{\rho_{ab}\otimes \rho_{cd}})\\
=&\sum_{iujvhk}s_i^2s_{u}^2r_j^2r_{v}^2\langle h_{b}k_{c}|U^{\dag}|u_b v_{c}\rangle\langle i_b j_c|U|h_{b}k_{c}\rangle\langle u_b v_c|U|h_{b}k_{c}\rangle\langle h_{b}k_{c}|U^{\dag}|i_b j_c\rangle\\
%=&\sum_{ef}(\sum_{ij}\lambda_i^2\mu_j^2\langle i_B j_C|U|e_{B}f_{C}\rangle\langle e_{B}f_{C}|U^{\dag}|i_B j_C\rangle)^{2}\\
=&\sum_{hk}(\langle h_{b}k_{c}|U^{\dag}\rho_{bc} U|h_{b}k_{c}\rangle)^{2}.
\end{aligned}
\end{equation*}
The nonbilocal measure for pure state is 
\begin{equation*}
\begin{aligned}
N_{\mathcal{A}}(|\Psi_{ab}\rangle\otimes|\Psi_{cd}\rangle)=& \max_{\Pi^{bc}} d_{\mathcal{A}}(\sqrt{\rho_{ab}\otimes\rho_{cd}},\Pi^{bc}(\sqrt{\rho_{ab}\otimes \rho_{cd}})) \\
=& 1-\min_{\Pi^{bc}} \mathcal{A}(\rho_{ab}\otimes \rho_{cd},\Pi^{bc}(\rho_{ab}\otimes \rho_{cd}))\\
%=&\max_{\Pi^{BC}} (\mathrm{tr}(\rho_{AB}\otimes \rho_{CD})^{2}-\mathrm{tr}(\rho_{AB}\otimes \rho_{CD})\Pi^{BC}(\rho_{AB}\otimes \rho_{CD}))\\
=&1-\min_{\Pi^{bc}}\sum_{hk}(\langle h_{b}k_{c}|U^{\dag}\rho^{bc} U|h_{b}k_{c}\rangle)^{2},
\end{aligned}
\end{equation*}
where the optimization is over all von Neumann measurements given in Eq. (\ref{measure}), leaving the marginal state $\rho^{bc}$ invariant. That is,
$$\rho^{bc}=\sum_{hk}\langle h_{b}k_{c}|U^{\dag}\rho^{bc}U|h_{b}k_{c}\rangle U|h_{b}k_{c}\rangle\langle h_{b}k_{c}|U^{\dag}$$
is a spectral decomposition of $\rho^{bc}$ since $\{U|h_{b}k_{c}\rangle\}$ is an orthonormal base. Comparing the above equation with Eq. (\ref{marginal}), we obtained
\begin{align}
N_{\mathcal{A}}(|\Psi_{ab}\rangle\otimes|\Psi_{cd}\rangle)=1-\sum_{i,j}s_i^4 r_j^4.
\end{align}
Hence the theorem is proved.
%\end{proof*}

\section*{Acknowledgment}
%\begin{acknowledgements}
This work was financially supported by the Council of Scientific and Industrial Research (CSIR), Government of India, under Grant No. 03(1444)/18/EMR-II.
%\end{acknowledgements}

\end{document}